\documentclass[aps,twocolumn,showpacs,amssymb,%
floatfix,superscriptaddress]{revtex4}
\usepackage{graphicx}
\usepackage{epsf}
\usepackage{dcolumn}
\usepackage{bm}
\usepackage{hyperref}
\usepackage{latexsym}
\usepackage{color}
\usepackage[croatian,english]{babel}
\usepackage{amssymb}
\usepackage{bm}
\usepackage{amsmath}

\newcommand{\beq}{\begin{equation}}
\newcommand{\eeq}{\end{equation}}

\newcommand{\arrows}{\leftrightarrow}
\begin{document}
\def\av#1{\langle#1\rangle}
\def\etal{{\it et al\/.}}
\def\pc{p_{\rm c}}
\def\l{{\lambda}}
\def\hm{h_*}
\def\xm{x_*}
\def\remark#1{{\bf *** #1 ***}}


\title{Model of Wikipedia growth based on information exchange via reciprocal arcs}

\author{Vinko Zlati\'{c}\thanks{vzlatic@irb.hr}}
\affiliation{Theoretical Physics Division, Rudjer Bo\v{s}kovi\'{c} Institute, P.O.Box 180, HR-10002 Zagreb, Croatia}
\affiliation{INFM-CNR Centro SMC Dipartimento di Fisica, Sapienza Universit\`a di Roma Piazzale Moro 5, 00185 Roma, Italy }
\author{Hrvoje \v{S}tefan\v{c}i\'{c}\thanks{shrvoje@thphys.irb.hr}}
\affiliation{Theoretical Physics Division, Rudjer Bo\v{s}kovi\'{c} Institute, P.O.Box 180, HR-10002 Zagreb, Croatia}

\begin{abstract}
We show how reciprocal arcs significantly influence the structural organization of Wikipedias, online encyclopedias. It is shown that random addition of reciprocal arcs in the static network cannot explain the observed reciprocity of Wikipedias. A model of Wikipedia growth based on preferential attachment and on information exchange via reciprocal arcs is presented. An excellent agreement between in-degree distributions of our model and real Wikipedia networks is achieved without fitting the distributions, but by merely extracting a small number of model parameters from the measurement of real networks.
\end{abstract}
\maketitle

\section{Introduction}

Since lately Wikipedias have been a vibrant interdisciplinary field of study ~\cite{Voss,NasWiki,Capocci,Muchnik,Huberman,Mehler,Giles}. The unique character of the free editing article policy and the large number of people participating in the process, make Wikipedia excellent model system for investigation of some complex system ideas in a realistic environment of the real social structure. Indeed, in the last few years there has appeared a growing amount of evidence supporting the usage of ideas from statistical physics, graph theory etc. in the description of the social or economic phenomena. This especially applies to phenomena which previously seemed untouchable from the natural scientists point of view~\cite{Palla,BarabasiQue,CaldaBanks}.

One of the very interesting features previously observed in Wikipedias is their reciprocity~\cite{NasWiki}. Reciprocal arcs are just the arcs pointing from the vertex $i$ to the vertex $j$ for which exists an arc pointing from vertex $j$ to the vertex $i$. The reciprocity is then defined as the fraction of reciprocal arcs in the total number of arcs $r=\frac{L^{\arrows}}{L}$~\cite{Reciprocity}. It was previously shown that reciprocal arcs can have an interesting role in real networks and in the theory describing them~\cite{WWW,Boguna,Zhou}. It also seems to be the most stable network measure one can find in the ensemble of Wikipedias except possibly the in-degree distribution exponent~\cite{NasWiki}. In~\cite{Gorka}, it was also shown that the reciprocity of Wikipedias cannot be explained by random mixing of arcs. In this paper we show in which manner reciprocal arcs influence the observed Wikipedia structure and show that they represent the necessary ingredient for understanding the Wikipedia growth and organization.

\section{Reciprocity in Wikipedia}

\begin{figure}[t]
\centering
\includegraphics*[width=0.5\textwidth]{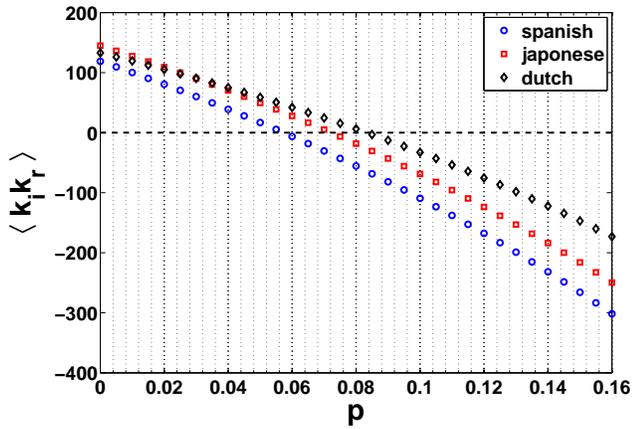}
\caption{\label{KiKrVsp.eps} The change of expected initial correlations of onevertex degrees $\langle k_ik_r \rangle$ calculated from the inverse of transformation matrix $\mathbf{T}^{-1}_{1v}$ for three different Wikipedias. Expected values of monitored correlations are changing the sign for the value of parameter about $p\sim 0.07$. Since the product moment correlations are strictly positive, this behavior indicates that there is just a small fraction of reciprocal arcs which are degree independent. In the case of Wikipedias the maximal value of parameter $p$ is about $p\sim 0.16$. }
\end{figure}

In order to understand how reciprocal arcs affect the structure of Wikipedia, first we need to examine if they exhibit any peculiar behavior at all.
 In the case of Wikipedia, one can expect existence of the reciprocal arcs between articles that share certain portion of content.
 If that were true, then the first assumption should be that the reciprocal arcs are distributed over the underlying Wikipedia network corresponding to mutual similarity of different articles.
 In other words, the content similarity of two articles is supposed to be independent of degrees of those two articles. One way to study this independence is laid out in the companion paper~\cite{NasTeorija}.
 There we show that the independence of reciprocal arcs can be studied using the the inverse matrix of the process of random addition of reciprocal arcs. Particularly we can use the equation
\abovedisplayskip=5 pt
\begin{equation}\label{JednadzbaInvTrans}
\langle \mathbf{S}(0)\rangle=\mathbf{T}_{1v}^{-1}(p)\mathbf{S}'(p),
\end{equation}
\belowdisplayskip=5 pt
to study such a process. In Eq. (\ref{JednadzbaInvTrans}) $\mathbf{S}'(p)$ represents the vector of product moments of degrees observed in the real network, $p=\frac{L^{\arrows}}{2L}$ represents the fraction of unidirectional arcs that were transformed to bidirectional, $\mathbf{T}_{1v}^{-1}$ is the inverse of the transformation matrix and $\langle \mathbf{S}(0)\rangle$ represents the expected vector of product moments of degrees in the network without reciprocal arcs. In Fig.~\ref{KiKrVsp.eps} we show that some types of correlations indicate that the assumption of the degree independent reciprocal arcs can not be justified in Wikipedia networks. It is obvious than the parameter $p$ can not be larger then $\sim0.07$ and from the data we know that the parameter $p$ should be around $\sim 0.16$.

This analysis is based on a very strong assumption of network stationarity. We know that the Wikipedia networks grow as many different Wikipedians edit many articles. Clearly it is necessary to investigate the influence of reciprocal arcs on the growth of Wikipedia.
In~\cite{NasWiki} we showed that different Wikipedias grow in a very similar fashion and that the number of newly added arcs is not linear with respect to the number of vertices. Nevertheless the observed behavior $L\sim N^{1.14}$ is close enough to linear that we can approximate it by a linear growth. 

\section{Model}

The model we use to describe the growth of Wikipedias is studied in detail in ~\cite{NasTeorija} and we just summarize the idea of the model and list the most important analytical results.
 The studied model was inspired by our findings in the Wikipedia networks~\cite{NasWiki}.
 Other authors studied the growth of Wikipedia networks with focus on preferential attachment~\cite{Capocci} and they found a linear-like relationship between the in--degree of the vertex and its probability to acquire a new arc, at least for the small and medium vertex degrees.
 It is also known that there is a significant portion of new arcs forming between old vertices in the network.
 Nevertheless this very often happens between vertices which are ``young'' compared to the age of the network~\cite{Personal}.
 This leads us to believe that ignoring additional formation of arcs betwen the older vertices is a reasonable approximation for the growth of the Wikipedia-like network.

The model consists of two steps. In the first one a new vertex, introduced in the network at time $t$ and therefore labeled as $t$, attaches to the network with $m$ outgoing arcs. The probability that the given arc, from these $m$ arcs, will attach itself to some vertex $s<t$ is proportional to the in--degree $k_i(s)$ of the vertex $s$. In the second step for every new arc with the probability $r$ a new reciprocal arc is formed between vertices $s$ and $t$. We showed that for such a model it is possible to find exact joint degrees probability distribution $P(k_i,k_o)$ of a single vertex. For example, in the case $m=1$, the distribution has the form
\begin{widetext}
\begin{equation}\label{konacnom1}
P(k_i,k_o)=\Theta(k_i-k_o)\binom{k_i-1}{k_o-1}r^{k_o-1}(1-r)^{k_i-k_o}\frac{r(1+r)}{2+r}\frac{(k_i-1)!}{(r+3)_{k_i-1}},
\end{equation}
\end{widetext}
 where $\Theta(x)$ is a usual Heaveside function and $(r+3)_{k_i-1}$ represents Pochammer symbol~\cite{Wolfram}. The asymptotic behavior of the marginal in-degree distribution for the described model is of the form:
\begin{equation}\label{KontRjesP(k_i)m1}
 P(k_i)\sim k_i^{-(2+r)}.
\end{equation}
This solution nicely interpolates between directed and undirected cases of the BA model~\cite{Usmjereni BA,BAJ99}. Furthermore, the asymptotic behavior of the in-degree distribution given in (\ref{KontRjesP(k_i)m1}) is also valid for any $m$, i.e. the power-law exponent does not depend on $m$.
 In~\cite{NasWiki} we reported the exponent of the in-degree distribution around $\gamma\simeq 2.18$  and values of reciprocity coefficient around $r_e\simeq0.35$. 
The described model for $m=1$ predicts the relations $r_e=2r/(1+r)$ and $\gamma = 2+r$, which explain the observed empirical values of $r_e$ and $\gamma$.

The three parameters which define the model are: $t$ - the size of the modeled network, $m$ - the number of outgoing arcs of the new vertex and $r$ - the probability of accompanying new arcs with their reciprocal arcs.
 In order to validate the model, we fixed three measured parameters of Wikipedia networks which uniquely describe the degree distributions obtained in the model.
 First the number of vertices in the monitored Wikipedia must be the same as the final size of the model network i.e. $t_{model}=N_{Wikipedia}$.
 In this way it is possible to check if the model also captures the details of the distribution in the tail as well as the power law exponent.
 The second parameter is the number of arcs in the modeled Wikipedia. 
The expected number of arcs obtained in the ensemble of model realizations has to be the same as the number of arcs measured in the modeled Wikipedia i.e. $E(L_{model})=L_{Wikipedia}$. The third parameter is the number of reciprocal arcs $L^{\arrows}_{Wikipedia}=E(L^{\arrows})$.
 The last two empirical parameters depend on our model parameters as
\begin{equation}\label{ExpNumbEdge}
L_{Wikipedia}=E(L_{model})=tm(1+r),
\end{equation}
and
\begin{equation}
 \label{expNumbRec}
L^{\arrows}_{Wikipedia}=E(L^{\arrows})=2tmr.
\end{equation}
From these equations it is easy to express our model parameters as functions of measured quantities: 
\begin{equation}
 \label{mForWiki}
m=\frac{L_{Wikipedia}-\frac{L^{\arrows}_{Wikipedia}}{2}}{N_{Wikipedia}},
\end{equation}
and
\begin{equation}
 \label{rForWiki}
r=\frac{L^{\arrows}_{Wikipedia}}{2L_{Wikipedia}-L^{\arrows}_{Wikipedia}}.
\end{equation}
\indent The parameter $m$ obtained from the measured quantities is not necessarily a natural number, which is supposed in our analytical treatment~\cite{NasTeorija}. In order to overcome this inconvenience we have used random numbers $\mathbf{m}$ drawn from Poisson distribution $E(\mathbf{m})=m$, to be the value of $m$ at any given time. We have shown that such a distribution has properties almost identical to our model with $m$ as a natural number if a suitable $E(\mathbf{m})$ is chosen~\cite{NasTeorija}. 

\section{Results}

\begin{figure}[t]
\centering
\includegraphics*[width=0.5\textwidth]{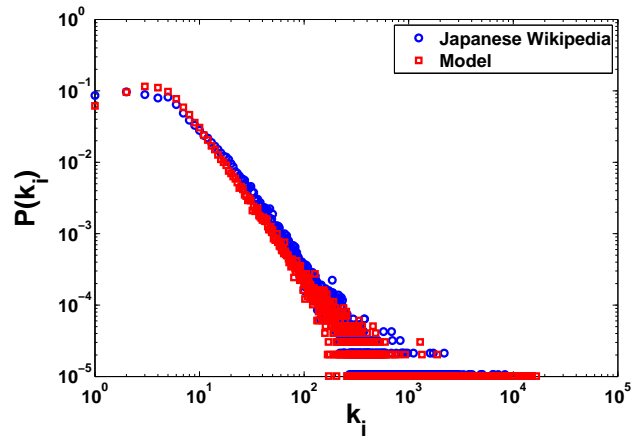}
\caption{\label{Fig: ModelVsWiki} Comparison of the in-degree distribution of the Japanese Wikipedia with one realization of our model for given parameters. It is easy to see excellent agreement both between mode of distribution, and exponent (slope in the log-log). Chosen parameters are $t=94094$, $m=16.75$, $r=0.18$. Our simulations show very similar behavior for the rest of studied Wikipedias. }
\end{figure}

In Fig. \ref{Fig: ModelVsWiki} we show an excellent agreement between the in-degree distribution of Japanese Wikipedia and our model.
 It is clear that the mode of the distribution is also well described with our model, a feature not so common in other degree distribution models found in the literature.
 We have already mentioned that in~\cite{Capocci} small and medium degrees show a kind of preferential attachment.
 For this reason it is important that our model describes well the mode which is formed by the vertices of relatively small in-degree.
 The tail of the distribution is also very well described by our model. This is important because such a tail was found to be a universal feature of Wikipedias in different languages. 

\begin{figure}[t]
\centering
\includegraphics*[width=0.5\textwidth]{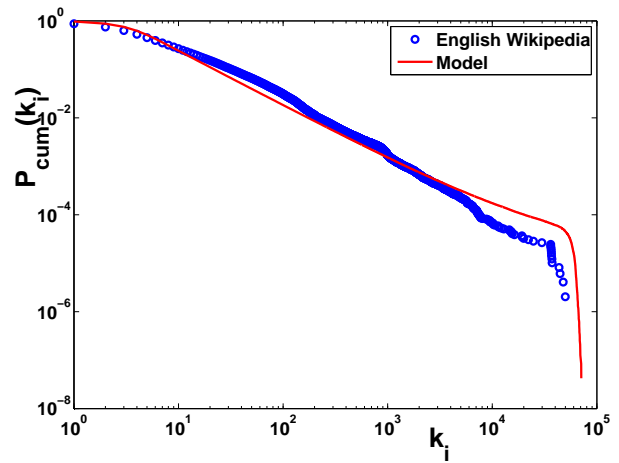}
\caption{\label{Fig: EnInModel} Comparison between cumulative in-degree distribution of English Wikipedia (blue circles) with one realization of our model (red line). Chosen parameters are $t=486291$, $m=18.24$, $r=0.15$. The distribution of the model follows closely the distribution of the model except in the very end of the tail, where we expect the aging effects in the real network.}
\end{figure}

If we compare a cumulative in-degree distribution of the model to the one of Wikipedias (see Fig. \ref{Fig: EnInModel}), it is again easy to see that our model shows a very good agreement in the tail of the distribution.
 One can also notice that our model shows a deviation from the monitored Wikipedia in the very end of the distribution.
 It is not surprising because we have already confirmed that the linearity in the attachment principle was observed only for small and medium degrees.
 The largest degrees show the ``aging'' effect i.e. they rarely attract new arcs, because their neighborhood is already matured in its content.
 We did not model such a behavior in order to keep the model as simple as possible.

\begin{figure}[t]
\centering
\includegraphics*[width=0.5\textwidth]{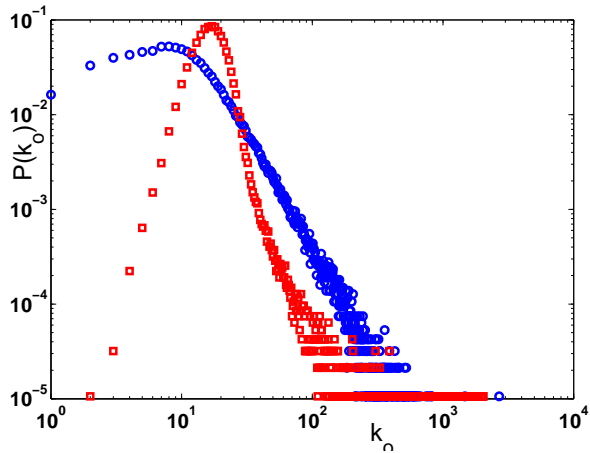}
\caption{\label{Fig: OutPrefRecip} Comparison between out-degree distribution of Japanese Wikipedia (blue circles) to one realization of our model (red squares). Chosen parameters are $t=94094$, $m=16.75$, $r=0.18$. There is no similarity between the modes of two distributions, and the slope of the tails seem to coincide in the very end of the distribution. }
\end{figure}

Our model does not reproduce out-degree distribution well (see Fig. \ref{Fig: OutPrefRecip}). The mode of the modeled distribution is shifted to much to the right and it is too narrow in comparison to the realistic out-degree distributions. This could be the consequence of using Poisson distribution for parameter $\mathbf{m}$, which is too narrow in this case. The reason we chose it is just because it is the most easily justifiable one parameter distribution for that case. Clearly we could get much better result with broader modal distributions for the parameter $\mathbf{m}$, but such a choice would be hard to justify and would be introduced just for fitting purposes. Since the aim of this paper is to clarify the fundamental role of the reciprocal arcs in the structure and growth of Wikipedia network, we focus on the version of the model which requires no fitting procedures. 

\section{Conclusion}

An excellent agreement of the Wikipedia and model in-degree distribution confirm that our model is a natural continuation of the process of preferential attachment, at least for the process of Wikipedia growth.
 Would the inclusion of the additional formation of new arcs between old vertices improve the agreement is discutable.
 The heuristics for additional changes in the model is not easy to justify without additional exploration of Wikipedia networks. 

It can be asserted that presented logic of Wikipedia growth can be attributed to the Wikipedians who are editing both old and new articles in a very small time frame.
 In such a case the reciprocity is also a good measure of the information interrelatedness in the knowledge networks.
 Clearly, the existence of the reciprocal arcs point to a certain intersection of the sets of information presented in different articles.
 Since reciprocity represents only the first viable correlation for such information sharing, it can be asserted that even better results could be expected if the model would take care of conservation of similar measures such as triad significance profile~\cite{Milo} or some other local structural motives.
 Taking into account the neighborhood of articles as a pool of more probable information sharing could also improve quality of the model.
 Problem with such attempts is the increase in the number of parameters which such models would require. 

The usefulness of our model in the case of networks of different origin is presently not clear.
 We feel that we have demonstrated a significant value of this model for understanding of Wikipedia networks and we believe that it could also be important in the case of other types of knowledge networks with time-dependent formation of arcs.
 Since reciprocity is a natural representation of feedback, the presented model and its extensions could aslo be useful in the study of complex systems in which feedback play an important role. 
The effort in this direction is a logical continuation of this research.

\section{Acknowledgments}

This work was financed by the Ministry of Education, Science and Sports of the Republic of Croatia under the contract No. 098-0352828-2863 and by the INFN, Italy. Vinko Zlati\'c is thankfull for support of G. Caldarelli and would like to thank A. Gabrielli for reading the manuscript.

\end{document}